\documentclass[12pt]{iopart}
\usepackage{epsfig}\input{epsf}
\usepackage{graphicx}

\usepackage{amssymb,amsfonts,amscd,latexsym}

\newcommand {\be} {\begin{equation}}
\newcommand {\ee} {\nonumber \end{equation}}
 \newcommand {\ber}{\begin{eqnarray*}}
 \newcommand {\eer} {\end{eqnarray*}}
\newcommand {\bea}{\begin{eqnarray}}
 \newcommand {\eea} {\end{eqnarray}}

\def \R {\mathbb{R}}

\def \bz{\bar z}

\def \ang {\frac{2\pi}{N}}

\newcommand{\apr}{\alpha'}

\newcommand{\de}{\delta}

\begin{document}

\begin{flushright}
CPHT-RR 014.0403\\
Imperial/TP/2-03/19\\
hep-th/0304015 \\
\end{flushright}

\title[D-branes in SL(2)/U(1)]{Semiclassical description of D-branes in SL(2)/U(1) gauged
WZW model}

\author{Angelos Fotopoulos }
\address{CPHT Ecole Polytechnique, Palaiseau 91128 France \\ and \\
 Theoretical Physics Group, Blackett Laboratory,
 Imperial College, London SW7 2BZ, U.K.}
\ead{angelos.fotopoulos@cpht.polytechnique.fr}

\begin{abstract}
In this note we examine some semiclassical features of D-branes in
the SL(2)/U(1) gauged WZW model and determine the small
fluctuation spectra for one class of branes. We compare our
results with expectations from the CFT side.
\end{abstract}

\section{Introduction}
It is well known that WZW models are some of the few known exact
string theory backgrounds. They are important elements of the
near-horizon geometry of various string theory configurations. For
example the background of $Q_5$ NS5 branes and $Q_1$ fundamental
strings has a near horizon geometry given by the $SL(2, \R) \times
SU(2)$ WZW model. On the other hand the near-horizon geometry of a
circular distribution of NS5 branes is T-dual to an orbifold of
the $SL(2, \R)/U(1) \times SU(2)/U(1)$ gauged WZW model. In
addition WZW models and their gauged variations have become
increasingly important in studies of the AdS/CFT correspondence
and branes-world scenarios.

D-branes are an essential ingredient of string theory since they
play crucial role in various dualities and holography. One way to
study the properties of D-branes in a given background is to
construct the boundary state of the D-branes which describe the
couplings of closed string modes on their world-volume. This is in
general a difficult question for a general background. Although
D-branes in WZW models on compact group manifolds like SU(2) have
been studied extensively, little progress has been achieved
towards non-compact cases like $SL(2, \R)$. In fact even the
spectrum of these theories is not fully understood. Nevertheless
we can get useful information for such D-branes by studying just
the semiclassical limit of their boundary states, in other words
their low energy effective action. This has been a fruitful
approach for the $SL(2, \R)$ case \cite{BP,PR,OT} as well as for
the cases of compact group manifolds \cite{BDS,BRS}. In this note
we will study the semiclassical features of D-branes in the gauged
$SL(2, \R)/U(1)$ WZW model along the lines of a similar study for
$SU(2)/U(1)$ \cite{MMS}. Since we are not in position to give the
full boundary state for all D-branes in $SL(2, \R)/U(1)$, we will
restrict ourselves to their geometrical features. Finally, for one
set of branes, we use a small fluctuation analysis of the low
energy effective action to get the semiclassical limit of the open
string spectra. Consequently we compare our results with
expectations from a CFT analysis \cite{RST}.

\section{The geometry and the spectrum}
The $SL(2, \R)$ WZW model is described by a world-sheet sigma
model with target space:
\begin{eqnarray} \label{SL2}
ds^2= k( -(\cosh \rho)^2 d \psi^2 + (d \rho)^2 + (\sinh \rho)^2 d
\tau^2)
\end{eqnarray}
 and a Kalb-Ramond field strength given by the volume element of the spacetime.
 The constant k is the level of the affine lie algebra that can
 take real values with $k \geq 2$, and we used units $\apr=1$.

One can construct the $SL(2, \R)/U(1)$ coset from the $SL(2, \R)$
WZW model by gauging a U(1) subgroup of $SL(2, \R)$
\cite{Wit,DVV}. If this subgroup is non compact one ends with a
manifold with Lorentzian signature, the 2D black hole. On the
contrary if we choose a compact U(1) then we get an Euclidean
manifold. There are two possible anomaly free U(1) gauge
symmetries: the axial, $g \to h g h $ and the vector, $g \to h g
h^{-1} $, where $g(z, \bz)$ group elements of $SL(2, \R)$ and $h
\in U(1)$. For a compact U(1) the manifold we get by gauging the
axial symmetry has the geometry of a cigar:
\begin{eqnarray}
& ds^2= (d \rho)^2 + (\tanh \rho)^2 d \tau^2 \qquad e^{-\Phi}=
e^{-\Phi_0} \cosh \rho \ \ (cigar)
\end{eqnarray}
The axial gauging amounts intuitively into gauging the symmetry
which corresponds to time translations in $SL(2, \R)$. This will
be important later in understanding the geometry of branes in the
cigar geometry. Notice that unlike $AdS_3$, in $SL(2, \R)$ we have
closed time-like curves. Therefore the U(1) associated with time
translations is a compact subgroup of $SL(2, \R)$.

By gauging the vector compact U(1) symmetry we get the geometry of
a trumpet:
\begin{eqnarray}
&ds^2= (d \rho)^2 + (\coth \rho)^2 d \psi^2 \qquad e^{-\Phi}=
e^{-\Phi_0} \sinh \rho \ \ (trumpet)
\end{eqnarray}
This coset amounts to gauging the U(1) symmetry of shifting
$\tau$. Notice that unlike the cigar manifold, that of the trumpet
is singular due to the fact that the vector gauging action has
fixed points. The two backgrounds are actually T-dual to each
other. To be exact the cigar is T-dual to the trumpet where $\psi
\sim \psi + 2 \pi/k$.

Finally we would like to summarize the operator spectrum of the
theory \cite{DVV} since it will become useful in our small
fluctuation analysis. The operators in the axially gauged theory
belong to discrete series with $(k-1)/2 \geq j \geq - 1/2$ and
continuous series with $j = -1/2 +iP$, $P \in R^+$. The operators
are charged under the $SL(2, \R)$ currents $J_3, \bar{J}_3$ with
charges: $\omega_L= 1/2(m+nk)$ and $\omega_R= - 1/2(m-nk)$. The
classical angular momentum of an operator is $m$ and the winding
$n$. The conformal weights of the operators are :
\begin{eqnarray}\label{cw}
h^j_{mn}= -\frac{j(j+1)}{k-2} +  \frac{(m+nk)^2}{k-2} \qquad
\bar{h}^j_{mn}= -\frac{j(j+1)}{k-2} +  \frac{(m-nk)^2}{k-2}
\end{eqnarray}
In addition from the harmonic analysis in $SL(2, \R)$ we conclude
that discrete representations should satisfy: $|nk|
> |m|$. We see that we cannot have discrete representations with
vanishing winding. For the T-dual vector gauging we need to take
$\omega_R \to - \omega_R$ which results in $m \leftrightarrow n$
in (\ref{cw}). The discrete representations constrain in this case
is $|nk| < |m|$, where $m$ is the angular momentum in the trumpet
and $n$ the winding. In this case discrete series states with
non-zero momentum and zero winding exist. The continuous
representations correspond semiclassically to wave like solutions
while the discrete ones to bound states. So, in the semiclassical
limit $k \to \infty$, where the winding states decouple, we expect
bound states for the trumpet and not for the cigar. These
expectations from CFT are nicely confirmed by the computation of
the low energy spectrum of the two geometries \cite{DVV}. The open
string spectra of the D-branes satisfy analogous constrains.

\section{D-brane descendants from branes of SL(2) and DBI analysis}
The geometry of D-branes in $SL(2,\R)$ has been studied in
\cite{BP} (see also i.e. \cite{Raj,Sar2,Que} for related work)
where it was found that solutions of the DBI action correspond to
regular and twined conjugacy classes of $SL(2, \R)$. It was found
that D1 branes have $AdS_2$ world volumes and D2 branes come in
two types with $H_2$ and $dS_2$ world volumes \cite{BP}. The first
type of D2 branes has a D-instanton density in its world-volume
while the second one was found to be unphysical due to a
supercritical electric field. As explained in
\cite{FS,Gaz,Sar,Wal} the D-branes in coset models $G/H$ have
world-volumes localized on the projection to the coset, of the
product of a conjugacy class of $G$ with a conjugacy class of $H$.
In geometric words we expect the D-branes we will find on the
cigar to be projections on a constant $\psi$ plane of the D-branes
in $SL(2, \R)$ and those of the trumpet projections on a constant
$\tau$ plane. We will verify these expectations by solving the DBI
for the cigar and trumpet.

We will assume from now on that the level $k$ is an integer. This
needs not to be the general case but in most of the interesting
(supersymmetric) backgrounds the $SL(2, \R)_k/U(1)$ CFT appears
with a corresponding $SU(2)_{k'}/U(1)$ with the levels $k$ and
$k'$ differing by an integer \cite{Koun}. Since $k'$ must be an
integer then $k$ should be one too.

For the D1 branes in the cigar and with an embedding $\rho(\tau)$
we need to minimize the "energy" of the system $ E =
\frac{\partial {\cal L}_{DBI}}{\partial \dot{\rho}} - {\cal
L}_{DBI}$, where $\dot{\rho}$ is derivative with respect to
$\tau$. The same method is employed for the D1 branes in the
trumpet. For the D2 branes one has to solve the equations of
motion for the gauge field $F_{\rho \tau}$. The analysis is
straight forward as shown in tables 1 and 2.

In tables 1 and 2 we give the embedding equations for the various
branes in trumpet and cigar respectively, as well as the $SL(2,
\R)$ branes from which they descent. The dimensionality of each
brane is evident by its name i.e. $DS1_{tr}$ is an 1-dimensional
brane in the trumpet, descending from the $dS_2$ brane in
$SL(2,\R)$.

\begin{eqnarray} \label{T1}
{\bf Table 1.} \nonumber \\
\begin{array}{|c||c|c|c|}
\hline SL(2,\R)  & {\rm Trumpet} & {\rm Embedding \ equations} &
{\rm Moduli}
\\
\hline H_2 & D1_{tr} & \cosh \rho \sin (\psi-\psi_0)=C & C=\sin
\ang k
\\
\hline dS_2 & DS1_{tr} & \cosh \rho \sin (\psi-\psi_0)=C & C\geq
1,\ \psi_0
\\
\hline AdS_2 & D2_{tr} & 2\pi
F_{\rho\psi}=\frac{N\coth\rho\sinh\rho_{min}}{
  \sqrt{\sinh^2\rho-\sinh^2\rho_{min}}} & \rho_{min},\ A_{\psi}
\\
\hline ? & D1'_{tr} & \rho=0 & none
\\
\hline
\end{array}
\nonumber
\end{eqnarray}

\begin{eqnarray} \label{T2}
{\bf Table 2.} \nonumber \\
\begin{array}{|c||c|c|c|}
\hline SL(2,\R) &{\rm Cigar} & {\rm Embedding\ equations} & {\rm
Moduli}

\\
\hline H_2 & D2_{cig}& 2\pi F_{\rho\tau}=N\frac{C\tanh
  \rho}{\sqrt{\cosh^2\rho-C^2}} & C=\sin \ang N_{D0}
\\
\hline dS_2 & DS2_{cig} & 2\pi F_{\rho\tau}=N\frac{C\tanh
  \rho}{\sqrt{\cosh^2\rho-C^2}} & C=\cosh \rho_{min}\geq 1,\ A_{\tau}
\\
\hline AdS_2 & D1_{cig}&  \sin (\tau -\tau_0)=\frac{\sinh
\rho_{min}}{\sinh \rho} & \rho_{min},\ \tau_0
\\
\hline ? & D0_{cig} & \rho=0 & none
\\
\hline
\end{array}
\nonumber
\end{eqnarray}

\begin{figure}
\begin{center}
\includegraphics[width=10cm]{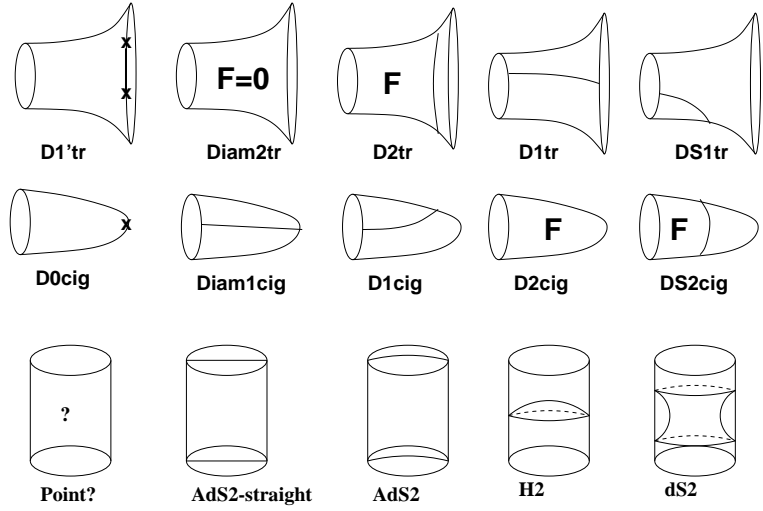}
\end{center}
\caption{The D-branes in the cigar and trumpet as well as the
D-branes of $SL(2, \R)$ where they descent from.
}\label{fig:branes}
\end{figure}


At this point a few comments are in order. First looking at the
table of the cigar and the trumpet we notice that the branes which
descent from the $AdS_2$ and $dS_2$ branes have two classical
moduli. Unlike these branes, the DBI analysis analogous to
\cite{MMS} for D2 branes in the cigar, implies that they carry a
D0 brane charge which leads to the quantization of the modulus
$C$. Also since they have trivial topology they do not admit a
Wilson line. By T-duality these branes are related to D1 branes in
the trumpet which hit the infinite radius circle at $\rho=0$. In
principal one expects no quantization of their incidence angle
given by the modulus $C$ and in addition that they carry an extra
modulus, the angle $\psi_0$ where they hit the circle $\rho=0$.
T-duality with the cigar suggests that these D1 branes can hit
only special points on the $\rho=0$ circle of the trumpet and it
is suggested also (see \cite{BFPR}) that they correspond to bound
states of D1 branes orthogonal to the circle with a number of
$D1'_{tr}$ branes.

We should also emphasize that the trumpet has a strong coupling
singularity at $\rho=0$ and therefore it is plausible that our
analysis might be valid only for branes which extend far enough
from the singularity. Nevertheless the considerable success of our
semiclassical analysis and the similar results of \cite{DVV} for
the spectrum of the theory suggest that there should be a grain of
truth in our results.

Notice that we have not been able to identify the brane of $SL(2,
\R)$, if any, where the $D0_{cig}$ branes on the cigar and their
T-dual $D1'_{tr}$ descent from. One might think that the D0 brane
of the cigar comes from the point brane of $SL(2, \R)$ but there
seems to be a contradiction when one tries to construct the
$D1'_{tr}$ by descending from this brane to the trumpet. The point
in $SL(2, \R)$ will still be a point in cigar or trumpet. These
shortcomings of our semiclassical analysis should be eliminated
once we are able to have the complete boundary state for these
branes.

Finally notice that since the cigar is actually T-dual to an
orbifold of the trumpet, one needs to rescale the string coupling
by $1/ \sqrt{N}$ when comparing quantities under T-duality.

\section{Small fluctuation analysis for $AdS_2$ descendants and the open string spectrum}
The boundary states for branes in $SL(2, \R)$ has been a long and
rather subtle problem. The non-compactness of the group results in
infinities for most physically relevant quantities which must be
dealt with caution. On top of that only recently \cite{MO} was a
significant aspect of the spectrum, that of spectral flow,
understood. Nevertheless one can work in the Euclidean analog of
$AdS_3$ and try to construct branes in the $H_3^+$ hyperbolic
space. This method was employed in \cite{PST} \footnote[1]{See
\cite{Ponsot} for a different approach to the problem and some
objections to the construction employed in \cite{PST}.} and the
boundary state of a brane with $AdS_2$ world-volume was
constructed among other branes of the background. It was claimed
that due to the time independence of this particular set of branes
on can descent trivially to the corresponding branes in the coset
$H_3^+ / \R_{\psi}$ which describes the cigar geometry.
 These branes couple only to closed string
momentum modes as expected by their geometry. This in turn
implies, due to the discrete representation constrains, that only
closed strings in the continuous representations couple to the
brane.

We will work along the lines of \cite{PST} to study the
semiclassical approximation to the open string spectrum of the
branes. It is easy to understand by the shape of the $D1_{cig}$
branes (see also \cite{OT}) that open string winding states should
appear in the spectrum and not momentum ones \footnote[2]{This is
also a consequence of closed/open string duality in the annulus
string amplitude. Momentum in the closed string channel becomes
winding in the open string channel.}. Since open string states
satisfy the same constrains for the allowed quantum numbers as
closed ones, continuous states as well as discrete ones with
winding $w > 0$ are allowed. In the semiclassical limit only the
winding zero state survives since all other ones become heavy and
decouple. Then computing the open string spectrum of these branes
using the DBI in the cigar geometry we expect to see only
continuous representations since the discrete ones are forbidden
for $w=0$. The purpose of the following analysis is to compute the
small fluctuation spectra on these branes and confirm the expected
spectrum. In addition we will compute the semiclassical reflection
coefficient and determine the density of open string states.

We use the DBI lagrangian for D1 strings with embedding $ \tau =
\cos^{-1} \frac{C}{r} + \de \tau (r,t)$ where $r= \sinh \rho, \ C=
\sinh \rho_{min}$ and we assume a fluctuation $\de \tau (r,t)$
around the classical background. The time is $t$ not to be
confused with $\tau$. Expanding the DBI action we get:
\begin{eqnarray}
&{\cal L}_{DBI} \simeq T_1 e^{-\Phi_0} [
\frac{1}{\sqrt{1-\frac{C^2}{r^2}}} + C(\partial_r \de \tau) \\
&-\frac{1}{2} k \frac{r^2 \sqrt{1-\frac{C^2}{r^2}}}{1+r^2}
(\partial_t \de \tau)^2 + \frac{1}{2} r^2
(\sqrt{1-\frac{C^2}{r^2}})^{3/2} (\partial_r \de \tau)^2]
\nonumber
\end{eqnarray}
The first term is the energy density of our D-brane. The first
order variation vanishes when we integrate it on the world-volume
of the brane since to have a small fluctuation $\de \tau \to
\tau_0$ for $x \to \pm \infty$ in cartesian coordinates.

The equations of motion for the quadratic fluctuations give a
hypergeometric equation. The solution shows no poles which would
signify bound states and it turns out that for the wave like
solutions with the correct asymptotic behavior the solutions is:
\begin{eqnarray}
&u=  \frac{\Gamma(-iP)}{\Gamma (1-iP/2)^2} r ^{-i P}
(1+C^2)^{iP/2} (-1)^{-i |j|/2} \\
&+ \frac{\Gamma(iP)}{\Gamma (1+iP/2)^2} r^{iP} (1+C^2)^{-iP/2}
(-1)^{+i |j|/2} \nonumber
\end{eqnarray}
where we have assumed (semiclassical approximation) $|j|
>>1$, for $j= -1/2 + iP$. The term $r^{-iP}$ describes an outgoing wave and $r^{iP}$ an incoming
one. The ratio of the coefficients of these two terms is the
semiclassical approximation to the reflection coefficient of a
state coming from infinity. Now, we can compute the semiclassical
limit of the relative density of states between a brane with
modulus $\rho_{min}$ and a reference brane $\rho_{min}^*$:
\begin{eqnarray}\label{PST}
N_{rel}(P|\rho_{min},\rho_{min}^*) \simeq \frac{1}{2 \pi i}
\frac{\partial}{\partial P} log
\frac{R_c(\rho_{min};P)}{R_c(\rho^*_{min};P)}=\frac{1}{\pi} log
\frac{\cosh \rho_{min}}{\cosh \rho_{min}^*}
\end{eqnarray}
where the function $R_c(\rho_{min};P)$ is the semiclassical limit
of the reflection coefficient for an open string state. We compute
a relative density of states since the density of states for a
single brane is infinite due to the non-compactness of its
world-volume. Our results agree with \cite{RST}.

\section{Conclusions }
In this note we have determined solutions of the DBI for D0, D1
and D2 branes in the cigar and trumpet geometries. Some aspects of
those branes in realistic string backgrounds will be further
discussed in \cite{BFPR}.We have also computed the small
fluctuations spectrum for one kind of them (the $D1_{cig}$) and
found agreement with geometrical and CFT expectations.

One might wonder if we can get useful information for the winding
states on D1cig, by studying the spectrum of momentum modes in its
T-dual, the $D2_{tr}$. We have calculated the small fluctuation
spectra of $D2_{tr}$ and found both discrete series states as well
as continuous representations. The discrete series states appear
with degeneracy one. For the continuous part of the spectrum the
density of states turns out to be the same as (\ref{PST}).

As a final remark, in \cite{OT} the spectral flow of semiclassical
solution gives strings which wind half a time around the center of
$AdS_3$. These in the cigar should descent to half-winding open
string states which is intuitively obvious from the shape of the
$D1_{cig}$. This implies that on the trumpet the $D2_{tr}$ should
allow states which have half-angular momentum and are periodic
only under a $4 \pi$ rotation. This deserves further investigation
and we hope to return to this point in the future.

\ack{ I wish to thank the organizers of the workshop on the
"Quantum Structure of Spacetime and the Geometric Nature of
Fundamental Interactions" held in Leuven from 13-19 September
2002, for giving me the opportunity to present this work in a very
stimulating environment. I am grateful to Costas Bachas, Marios
Petropoulos and Sylvain Ribault for many discussions and
collaboration in this and related subjects as well as to the
authors of \cite{RST} for communicating their results. I
acknowledge discussions with A. Giveon, V. Schomerus, A. Schwimmer
and J. Teschner. Finally I wish to thank J. Troost for bringing to
my attention a mistake in a formula of the first version. This
work was supported in part by the European Commission under
contracts HRPN-CT-2000-00131 and HRPN-CT-2000-00148.}

\end{document}